\documentstyle[aps]{revtex}
\newcommand{\partl}[2]{\frac{\partial #1}{\partial #2}}
\newcommand{\xb}{{\bf x}}
\newcommand{\kb}{{\bf k}}
\newcommand{\yb}{{\bf y}}

\begin{document}
\include{psfig}
\bibliographystyle{srt} 
\title{\bf Rigorous estimates of the tails of the 
probability distribution function for the random linear shear model.}
\vspace{-.1in} 
\author{Jared C. Bronski}
\address{Department of Mathematics, University of Illinois Urbana Champaign,
Urbana, IL 61820.}
\author{Richard M. McLaughlin}
\address{Department of Mathematics, University of North Carolina,
Chapel Hill, NC 27599.}
\author{}\vspace{-.15in}

\address { In previous work Majda\cite{Majda1,Majda2} and 
McLaughlin\cite{McLMa,Rico} 
computed explicit expressions for the $2N$th moments of a passive scalar 
advected by a linear shear flow in the form of an integral over ${\bf R}^N$. 
In this paper we first compute the asymptotics of these moments for large 
moment number. We are able to use this information about the large 
$N$ behavior of the moments, along with some basic facts about  
entire functions of finite order, to compute the asymptotics of the tails 
of the probability distribution function. We find that the probability 
distribution has Gaussian tails when the energy is concentrated in 
the largest scales. As the initial energy is moved to smaller and smaller 
scales we find that the tails of the distribution grow longer, and the
distribution moves smoothly from Gaussian through exponential and 
``stretched exponential''. We also show that the derivatives of the scalar 
are increasingly intermittent, in agreement with experimental observations, 
and relate the exponents of the scalar derivative to the exponents of the 
scalar.}

\maketitle

\section{Background}

It is a well documented experimental fact that, while the statistics 
of the velocity field in a turbulent flow are roughly Gaussian, the statistics 
of other quantities like the pressure, derivatives of velocity and
a passively advected scalar are generally far from 
Gaussian.\cite{CGH,CGHKLTWZZ,Ching,GCGLN,KSS,ThVanA} For example Castaing, et. 
al.\cite{CGHKLTWZZ} observed in experiments
in a Rayleigh-B\'enard convection cell that for Rayleigh number $Ra<10^7$ 
the distribution of temperature appeared to be roughly Gaussian, while for 
larger Rayleigh numbers, $Ra>10^8$, the temperature distribution appeared 
to be closer to exponential. In related work Ching\cite{Ching} studied the 
probability distribution functions (pdfs) for temperature differences at 
different scales, again in a Rayleigh-B\'enard cell, and found that the pdfs 
over a wide range of scales were well approximated by a `stretched exponential'
distributions of the form 
\[
P(T) = e^{-C |T|^\beta}.
\]
At the smallest scales the observed value of the exponent was 
$\beta\approx .5$, while at the largest scales the observed exponent 
was roughly $\beta \approx 1.7$. Kailasnath, Sreenivasan and 
Stolovitky\cite{KSS} measured the pdfs of 
velocity differences in the atmosphere 
for a wide range of separation scales. They found similar distributions to 
the ones found by Ching, with exponents ranging from $\beta\approx.5$ for 
separation distances in the dissipative range to $\beta \approx 2$ on the 
integral scale. Finally Thoroddsen and Van Atta\cite{ThVanA} studied 
thermally stratified turbulence in a wind tunnel and found the 
probability distributions of the density to be roughly Gaussian, while 
the distributions of the density gradients were exponential.   

A complete understanding of such intermittency lies at the heart 
of understanding fluid turbulence, and would certainly require 
a detailed understanding of the creation of small scale fluid 
structures involving both patchy regions of strong vorticity and 
intense gradients \cite{Chorin,Sreenivasan2}.  An alternative 
starting point is to assume the statistics of the flow are 
known a priori and to determine how these statistics are manifest in 
a passively evolving quantity.  This question of inherited statistics 
is significantly easier than the derivation of a complete theory for 
fluid turbulence, though still retains many inherent difficulties 
such as problems of closure.  

Motivated by the Chicago experiments of the late 80's \cite{CGHKLTWZZ}, and 
earlier work\cite{AS,LL,PS,Sre2}, there has been 
a tremendous effort towards understand the origin of the intermittent 
temperature probability distribution function in passive scalar 
models with prescribed (usually Gaussian) velocity statistics.  For a 
very complete review of the subject of turbulent diffusion, including 
a full discussion of scalar intermittency, see 
the recent survey article of Majda and Kramer \cite{MajdaKramer}.  
Most of the work on the scalar statistics has either been directed at 
understanding the anomalous scaling of temperature structure functions, 
or at understanding the shape of the tail of the limiting scalar pdf.  

There has been a wealth of theoretical efforts addressing this last 
issue 
of the tail\cite{BaFa,Us,CGHKLTWZZ,CCK,Ch,CFKL,ChiTu,Deu,HoSig1,pierrehumbert,K,Kerstein,Majda1,Majda2,McLMa,Pumir,PumirShraimanSiggia,ShrSig,SiYak,Son,YOBJSS}. 
A somewhat common theme, particularly in the pumped 
case, is the prediction that the scalar pdf should 
develop an exponential tail.  For example 
Kraichnan\cite{K}, Shraiman and Siggia\cite{ShrSig} 
and Balkovsky and Falkovich\cite{BaFa} all find exponential tails.  
Another important question is to understand the pdf of the 
scalar gradient. Naturally, gradient information may 
be expected to amplify contributions from small scales, and a general theory 
relating the scalar tail with the gradient tail, even for passively evolving 
quantities would be quite valuable. There has been somewhat less theoretical 
effort aimed at exploring the difference in statistics between 
the scalar and the scalar gradient. Chertkov, Falkovich and 
Kolokolov\cite{CFK}, Chertkov Kolokolov and Vergassola\cite{CKV} and 
Balkovsky and Falkovich\cite{BaFa}
have explored this question and have found a stretched 
exponential distribution of the scalar gradient in situations 
for which the scalar has an exponential tail. Holzer and 
Siggia\cite{HoSig1,HoSig2}, and Chen and Kraichnan\cite{CK} have 
observed similar phenomena numerically.  

In this paper we examine the scalar and scalar gradient pdf tail in an exactly 
solvable model first studied by Majda \cite{Majda1} and 
McLaughlin and Majda \cite{McLMa} who were able to construct explicit 
moment formulas for the moments of a passive scalar advected by a rapidly 
fluctuating linear shear flow in terms of $N$-dimensional integrals. 
In that work, it 
was established that the degree of length 
scale separation between the initial scalar field and the fluid flow is 
inherent to the development of a broader than Gaussian pdf.  

Here, we explicitly calculate the tails of the pdf for this model.  
We begin by analyzing the expression derived by Majda 
for the large time $2N$th moment of the pdf for the random uniform 
shear model, which is given by an integral over ${\bf R}^N$.   From these 
normalized moments, we will construct the tail of the associated pdf.  
We point out that in this calculation the convergence of the 
pdf for finite time to the pdf for infinite time is weak - for 
fixed moment number the finite time moment converges to the 
limiting moment. The convergence is almost certainly not uniform 
in the moments. 
For a more thorough investigation of the uniformity of this 
limiting process in the context of general, bounded 
periodic shear layers, see Bronski and McLaughlin\cite{Us}.

The tail is calculated in two steps.  First, 
using direct calculation and gamma function identities we are able 
to reduce the $N$-dimensional integral to a {\em single} integral 
of Laplace type, from which 
the asymptotic behavior of the $2N$th moment follows easily. The asymptotic
behavior of the moments is important for determining the tails of the 
probability distribution function, as we establish below.  
Second, we consider the problem of reconstructing 
the probability measure from the moments. 
Using ideas from complex analysis, mainly some basic facts about 
entire functions of finite order and type, we are able to provide 
rigorous estimates for the rate 
of decay of the tails of the measure. We find that the tails 
decay like
\[
\exp(-c_\alpha|T|^{\frac{4}{3+\alpha}})
\]
so depending on the precise value of the parameter $\alpha$ 
(defined in section II, below, 
which sets the degree of scale separation between 
the scalar and flow field) the model admits tails which 
are Gaussian, exponential, or stretched exponential. We also show that in this 
model higher order derivatives of the scalar in the shear 
direction are always more intermittent, with a very simple relationship 
between the exponents of the scalar and its derivative. The distributions 
of derivatives in the cross-shear direction, however, display the same 
tails as the scalar itself.  

We remark that, while the stream-line topology for shear profiles 
is admittedly much simpler than that in fully developed turbulence, 
the fact that the exact limiting tail for the decaying scalar 
field may be explicitly and rigorously constructed suggests such 
models to be exceptionally attractive for testing the validity of 
different perturbation schemes. It is also extremely interesting 
because it demonstrates that, at least for unbounded flows, a positive 
Lyapunov exponent (as would typically occur for a general Batchelor flow) 
is {\em not necessary} for intermittency. For an interesting
discussion of the role of Lyapunov exponents in producing intermittency 
see the work of Chertkov, Falkovich, Kolokolov and Lebedev.\cite{CFKL} 

\subsection{The random shear model}
Here, we briefly review the framework of the random shear 
model\cite{Majda1,Majda2,McLMa,Us}.  
We follow Majda, and consider the free evolution of a passive 
scalar field in the presence of a rapidly fluctuating 
shear profile:
\begin{eqnarray}
\label{scalar}
\partl{T}{t} + \gamma(t) v(x) \frac{\partial T}{\partial y} 
&=& \bar \kappa \Delta T.
\end{eqnarray}
The random function, $\gamma(t)$, represents multiplicative, 
mean zero Gaussian white noise, delta correlated in time:
\begin{eqnarray*} 
\left<\gamma(t) \gamma(s)\right> = \delta(|t-s|)
\end{eqnarray*}
where the brackets, $\left<\cdot\right>$, denote the ensemble average 
over the statistics of $\gamma(t)$.  
The original model considered by Majda involved the case of 
a uniform shear layer, $v(x)=x$, which leads to the moments considered 
below \cite{Majda1}.  It a quite general fact, not special to 
shear profiles, that a closed evolution equation for the arbitrary 
N-point correlator is available for the special case of rapidly 
fluctuating Gaussian noise, see work of Majda \cite{Majda2} for 
a path integral representation of this fact for the special 
case of random shear layers.  For the scalar evolving in 
(\ref{scalar}), the N point correlator, defined as:
\begin{eqnarray}
\label{correlator}
\psi_N(\xb,\yb,t) &=& \left<\prod_{j=1}^N T(x_j,y_j,t)\right>\\
\xb &=& (x_1,x_2,x_3,...,x_N) \nonumber \\
\yb &=& (y_1,y_2,y_3,...,y_N) \nonumber
\end{eqnarray}
is a function:  $\psi_N : R^{2N}\times [0,\infty) \rightarrow R^1$ satisfying
\begin{eqnarray}
\label{corevolve}
\partl{\psi_N}{t} &=& \bar \kappa \Delta_{2N} \psi_N + 
\frac{1}{2} \sum_{i,j=1}^N v(x_i)v(x_j) \frac{\partial^2 \psi_N}
{\partial y_i \partial y_j}
\end{eqnarray}
where $\Delta_{2N}$ denotes the $2N$ dimensional Laplacian.  

We next describe the initial scalar field.  Following Majda\cite{Majda1}, 
we assume that the scalar is initially a mean zero, Gaussian random 
function depending only upon the variable, $y$: 
\begin{equation}
\label{data}
T|_{t=0} = \int_{R^1} dW(k) e^{2 \pi i k y} |k|^{\frac{\alpha}{2}} 
\hat \phi_0(k)\qquad \alpha > -1 
\end{equation}
Here, $\hat \phi_0(k)$ denotes a rapidly decaying (large k) cut-off 
function satisfying $\hat\phi_0(k)=\hat\phi(-k), \hat\phi_0(0)\ne 0$
and $dW$ denotes complex Gaussian white noise with 
\begin{eqnarray*}
\left<dW\right>_W &=& 0\\
\left<dW(k)dW(\eta)\right>_W &=& \delta(k+ \eta) dk d\eta
\end{eqnarray*}

The spectral parameter, 
$\alpha$ appearing in (\ref{data}) is introduced to adjust the excited 
length scales of the initial scalar field, with increasing $\alpha$ 
corresponding to initial data varying on smaller scales.  
We remark that the more general case involving initial data depending upon both 
$x$ and $y$, and data possessing both mean and fluctuating components, 
was analyzed McLaughlin and Majda \cite{McLMa}.  

For this case 
involving shear flows, the evolution of this $N$ point correlator may be 
immediately converted to parabolic quantum mechanics through 
partial Fourier transformation in the $\yb$ variable. For the particular 
initial data presented in (\ref{data}), this yields the following solution 
formula:
\begin{eqnarray*}
\psi_N = \int_{R^N} e^{2 \pi i \kb\cdot \yb} \hat \psi_N(\xb,\kb,t) 
\prod_{j=1}^N \hat \phi_0(k_j) |k_j|^{\frac{\alpha}{2}}dW(k_j)
\end{eqnarray*}
where the N-body wavefunction, $\hat \psi_N(\xb,\kb,t)$ satisfies the 
following parabolic Schr\"odinger equation:
\begin{eqnarray}
\label{schrod}
\partl{\hat \psi_N}{t} &=& \bar \kappa \Delta_{\xb} - V_{int}(\kb,\xb)
\hat \psi_N\\
\hat \psi_N|_{t=0}&=&1 \nonumber
\end{eqnarray}
The interaction potential, $V_{int}(\kb,\xb)$, is 
\begin{eqnarray*}
V_{int}&=& 4 \pi^2 |\kb|^2 +2 \pi^2 (\sum_{j=1}^N k_j v(x_j))^2 .
\end{eqnarray*}

For the special case of a uniform, linear shear profile, with 
$v(x)=x$, the quantum mechanics problem in (\ref{schrod}) is exactly 
solvable in any spatial dimension.  Taking the ensemble average 
over the initial Gaussian random measure using a standard cluster 
expansion, the general solution formula for $\left<\psi_N(\xb,\yb,t)\right>_W$ 
is obtained \cite{Majda1,McLMa} in terms of $N$ dimensional integrals.  
The normalized, long time flatness factors, $\mu^\alpha_{2N}=
\lim_{t\rightarrow \infty}\frac{\left<T^{2N}\right>}
{\left<T^2\right>^N}$, are calculated by 
evaluating the correlator along the diagonal, 
\begin{eqnarray*}
\xb &=& (x,x,x,\cdots,x)\\
\yb &=& (y,y,y,\cdots,y)
\end{eqnarray*}
and utilizing the explicit long time asymptotics available through 
Mehler's formula.  This leads to the following set of normalized 
 moments for the decaying scalar field, $T$:
\begin{eqnarray}
\mu^\alpha_{2N}&=& 
\frac{(2N)!}{2^N N! \sigma^N} \int_{R^N} d\kb \frac{\prod_{j=1}^N 
|k_j|^{\alpha}}{\sqrt{\cosh(|\kb|)}}\\
\sigma &=& \int_{R^1} dk \frac{|k|^{\alpha} }{\sqrt{\cosh{|k|}}}. \nonumber
\end{eqnarray}

Observe that these normalized moments depend upon the parameter 
$\alpha$.  By varying this parameter Majda and McLaughlin established 
that the degree of scale separation between the initial scalar and flow field 
is important in the development of a broader than 
Gaussian pdf \cite{Majda1,McLMa}.  
They demonstrated through numerical quadrature of these integrals 
for low order moments that as the initial scalar field develops an 
infrared divergence (with $\alpha\rightarrow -1$, corresponding to the 
loss of scale separation between the initial scalar field, and the 
infinitely correlated linear shear profile) the limiting single point scalar 
distribution has Gaussian moments\cite{Majda1}.  Conversely they showed
that as the length scale of the initial scalar field is reduced, corresponding
to increasing values of $\alpha$, the limiting distribution shows growing 
moments indicative of a broad tailed distribution\cite{McLMa}.  
On the basis of these low order moment comparisons, 
these studies suggest that within these models, the 
limiting pdf should be dependent upon the scale 
separation between the scalar and flow field.  A fundamental 
issue concerns whether and how this scale dependence is 
manifest in the pdf {\em tail}.  Below, we address precisely this 
issue, and rigorously establish that the intuition put 
forth by Majda and McLaughlin is correct 
through the explicit calculation of the limiting pdf tail.

\section{Asymptotics of the probability distribution}

\subsection{Notation}
Recall from the previous section that the work of Majda derived exact 
expressions for the moments of a one parameter family of models indexed 
by the exponent $\alpha$.  
In the remainder of the paper $d\mu^\alpha(T)$ will denote the probability 
measure for the passive scalar $T$ in the Majda model with exponent $\alpha$. 
The $i^{th}$ moment of the probability measure $d\mu^\alpha(T)$ will be 
denoted by $\mu^\alpha_i$. In this particular model the distribution is 
symmetric and thus all odd moments vanish. 
   
\subsection{Large $N$ asymptotics of the moments}

In this model the exact expression for the $2N$th
moment is given by
\begin{eqnarray*}
\mu^\alpha_{2N} &=& {\frac{(2N)!}{\sigma^N 2^N N!}} 
\int \frac{\prod_{j=1}^N |k_j|^\alpha}
{\sqrt{\cosh(|\vec k|)}} dk_1 dk_2 dk_3\dots dk_N \\
\sigma &=& \int \frac{|k|^\alpha d\!k}{\sqrt{\cosh(k)}}  
\end{eqnarray*}
As noted by Majda $\cosh(|\vec k|)\le \prod \cosh(k_i)$ which implies 
the normalized flatness factors are strictly larger than those of a 
Gaussian, implying broad tails. 
The simplest way to analyze this integral, and in particular to understand 
the behavior for large $N$, is to introduce spherical coordinates. 
Spherical coordinates in $N$ dimensions 
can easily be constructed iteratively in terms of spherical coordinates 
in $N-1$ dimensions as follows. The coordinates in $N$ dimensional 
spherical coordinates are $\{r,\theta_1,\theta_2,\theta_3\dots\theta_{N-1}\}.$
If $\{x^{N-1}_1,x^{N-1}_2\dots x^{N-1}_{N-1}\}$ are coordinates on 
${\bf R}^{N-1}$ then coordinates on ${\bf R}^N$ are given by 
\begin{eqnarray*}
x^N_i &=& x^{N-1}_j \sin(\theta_{N-1}) \qquad j \in 1\dots N-1 \\
x^N_N &=& r \cos{\theta_{N-1}} 
\end{eqnarray*}
Using this construction it is simple to calculate that the volume element 
in $N$ dimensional spherical coordinates is given by 
\[
dx_1 dx_2 \dots dx_N = r^{N-1} dr \prod_{j=1}^{N-1} \sin^{j-1}(\theta_j) 
d\theta_j \qquad \theta_1 \in [0,2\pi] \qquad \theta_{i>1} \in [0,\pi]. 
\]
Since the volume element is a product measure the $N$ dimensional integral
factors as a product of $N$ one dimensional integrals and we are left with the 
expression 
\[
\mu^\alpha_{2N} = {\frac{(2N)!}{\sigma^N 2^N N!}} I_0(N) 
\prod_{j=1}^{N-1} I_j,
\]
where the $I_j$ are given by 
\begin{eqnarray}
I_0(N) &=& \int_0^\infty r^{N(\alpha+1)-1} {\frac{dr}{\sqrt{\cosh(r)}}} \nonumber \\ 
I_1  &=& \int_0^{2\pi} |\sin(\theta)|^{\alpha} 
|\cos(\theta)|^{\alpha} d\theta \nonumber \\
I_j &=& \int_0^\pi |\sin(\theta)|^{j(\alpha + 1)-1} 
|\cos(\theta)|^{\alpha} d\theta \qquad j > 1. 
\end{eqnarray}

The angular integrals can be done explicitly in terms of gamma functions, 
using the beta function identity
\[
2\int_0^{\pi/2} |\sin(\theta)|^{2z-1}|\cos(\theta)|^{2w-1} d\theta =
\beta(z,w) =  
{\frac{\Gamma(z)\Gamma(w)}{\Gamma(z+w)}}
\]
which leads to the expression 
\begin{eqnarray}
\mu^\alpha_{2n} &=& 2{\frac{(2N)!}{\sigma^N 2^N N!}} I_0(N)
\prod_{j=1}^{N-1} \frac{\Gamma(\frac{\alpha+1}{2})\Gamma(j\frac{\alpha+1}{2})}
{\Gamma((j+1)\frac{\alpha+1}{2})} \nonumber\\ 
&=&  2{\frac{(2N)!(\Gamma(\frac{\alpha+1}{2}))^{N-1}}{\sigma^N 2^N N!}} I_0(N)
\prod_{j=1}^{N-1} \frac{ \Gamma(j\frac{\alpha+1}{2})}{\Gamma((j+1)\frac{\alpha+1}{2})}.
\end{eqnarray}
Observe that the product telescopes - the numerator of one term is the denominator of the next - giving the final expression
\begin{eqnarray}
\mu^\alpha_{2n} &=& 2{\frac{(2N)!}{\sigma^N 2^N N!}}
\frac{(\Gamma(\frac{\alpha+1}{2}))^N}{\Gamma(N\frac{\alpha+1}{2})} 
\int r^{N(\alpha+1)-1} {\frac{dr}{\sqrt{\cosh(r)}}} \nonumber\\
&=&  2{\frac{(2N)!}{\sigma^N 2^N N!}}
\frac{(\Gamma(\frac{\alpha+1}{2}))^N}{\Gamma(N\frac{\alpha+1}{2})} I_0(N)
\label{eqn:mu2n} 
\end{eqnarray}
The integral over the radial variable $I_0(N)$ 
cannot be done explicitly, but the
large $N$ asymptotics are given by 
\[
I_0(N) \approx 2^{N(\alpha+1)+\frac{1}{2}}\Gamma(N(\alpha+1)), 
\]
so that the large $N$ behavior of the moments is given by 
\begin{equation}
\mu^\alpha_{2N} \approx 2^{N\alpha + \frac{3}{2}}{\frac{(2N)!}{\sigma^N N!}}
\frac{\Gamma(N(\alpha+1))(\Gamma(\frac{\alpha+1}{2}))^N
}{\Gamma(N(\frac{\alpha+1}{2}))}. 
\label{eqn:moment_asymp}
\end{equation}
Note that since 
\[
\frac{\Gamma(N(\alpha+1))}{\Gamma(\frac{N(\alpha+1)}{2})} \rightarrow \infty \quad 
{\rm as\; }N \rightarrow \infty
\]
the moments are strictly larger than the moments of the Gaussian. We 
will use this to provide rigorous quantitative estimates 
for the tails of the distribution.  

\subsection{The Hamburger moment problem}
Having computed simple expressions for the moments of the pdf, 
as well as asymptotic expressions for large moment number, 
it is natural to ask the question of whether one can do the 
inverse problem, and deduce the pdf itself. 
The problem of determining a measure from its moments is a 
classical one, known as the Hamburger moment problem\cite{RS,ShTa,Wi}.
This problem has a rich theory, and we mention only a very few of the most
basic results here. For an overview of the subject, see the book by Shohat 
and Tamarkin\cite{ShTa} or the recent electronic preprint by Simon\cite{BS}.
  
The two most important questions are, of course, existence and uniqueness. 
There is a necessary and sufficient condition for a set of 
numbers $\{ \mu_i\}$ to be the moments of some probability 
measure, namely that the expectation of any positive polynomial 
be positive. This translates into the following linear algebraic 
conditions on the diagonal determinants of the Hankel matrix, the matrix 
with $i,j^{th}$ entry $\mu_{i+j}$:
\[
\left| \mu_0 \right| > 0, \qquad 
\left| \begin{array}{cc}
\mu_0 & \mu_1 \\
\mu_1 & \mu_2 
\end{array}\right|  > 0,
\qquad 
\left|\begin{array}{ccc}
\mu_0 & \mu_1 & \mu_2 \\
\mu_1 & \mu_2 & \mu_3 \\
\mu_2 & \mu_3 & \mu_4 
\end{array} \right| > 0 \ldots 
\]
These conditions appear to be quite difficult to check in practice. However 
since the moments considered here are, by construction, the moments of a pdf 
this condition must hold. 

A more subtle question is the issue of uniqueness of the measure, usually 
called determinacy in the literature of the moment problem.  
One classical sufficient condition for the determinacy of the moment problem 
is the following condition, due to Carleman\cite{Ca,ShTa}:
If the moments $\mu_n$ are such that the following sum {\em diverges} 
\[
\sum_{j=1}^\infty (\mu_{2j})^{-\frac{1}{2j}} = \infty 
\]
then the Hamburger moment problem is determinate. 
Given the asymptotic expression for the moments given in Equation 
(\ref{eqn:moment_asymp}) it is easy to check that
\[
(\mu_{2j}^\alpha)^{-\frac{1}{2j}} \approx c j^{-\frac{\alpha+3}{4}}
\]
and thus there is a unique measure with these moments 
for $-1 \le \alpha \le 1$.  We will see later that this corresponds to 
probability distribution functions with tails that range from Gaussian 
through exponential. 
 
In the case $\alpha > 1$ which, as we will see later, corresponds to 
stretched exponential tails, the problem 
probably does not have a unique solution. Indeed there are 
classical examples of collections of moments with the same asymptotic 
behavior as the stretched exponential distribution for which the moment 
problem has a whole family of solutions. 

Given this we come to the question of actually calculating the 
measure given the moments. There is a rather involved theory for this 
in the determinate case involving, among other things, orthogonal 
polynomials and continued fractions\cite{KrMcL,ShTa}, but in general 
this problem is extremely difficult. However we show in the next section 
that it is relatively straightforward to reconstruct the 
{\em tails of the measure} from the moments.  
 
\subsection{Asymptotics of the tails of the distribution}

Recall that $\mu^\alpha_{2N}$ is the $2N$th moment of some probability measure 
$d\!\mu^\alpha(T)$,
\begin{equation}
\mu_{2N}^\alpha = \int T^{2N} d\!\mu^\alpha(T).
\end{equation}
We are interested in calculating the asymptotic rate of decay  of the tails 
of the probability measure $d\!\mu^\alpha(T)$. The information about the 
behavior of the tails of the distribution is contained in the asymptotic 
behavior of the large moments. We study the tails of the measure 
$d\!\mu^\alpha(T)$ by introducing the function 
\begin{equation}
f^\alpha(z) = \sum_{j=0}^\infty \frac{\mu_{2j}^\alpha z^{2j}}
{\Gamma(\frac{j(3+\alpha)}{2})C^{2j}},  
\end{equation}
where $C$ is some as yet unspecified constant. 
The factor of $\Gamma(\frac{j(3+\alpha)}{2})$ is chosen so that the series 
for $f^\alpha$ has a finite but non-zero radius of convergence. This will 
give us the sharpest control over the tails of $d\!\mu^\alpha(T)$. 
It is convenient to demand that the radius of convergence of the series 
be one. Using the root test it is straightforward to check that the radius 
of convergence of the sum is given by 
\[
r^*=  
C 2^{-(\alpha+2)}\frac{(\alpha+3)^{\frac{\alpha+3}{4}}}
{ (\alpha+1)^{\frac{\alpha+1}{4}}}
\sqrt{\frac{\sigma}{\Gamma(\frac{\alpha+1}{2})}},
\]
so we choose 
\[
C = 2^{\alpha + 2} 
\frac{(\alpha+1)^{\frac{\alpha+1}{4}}}{(\alpha+3)^{\frac{\alpha+3}{4}}} 
\sqrt{\frac{\Gamma(\frac{\alpha+1}{2})}{\sigma}}.
\]

Since the coefficients $\mu_{2N}^\alpha$ 
are the moments of a probability measure $d\!\mu^\alpha(T)$ 
we have the alternative expression
\[
f^\alpha(z) = \sum_{j=0}^\infty \frac{z^{2j}}{C^{2j} 
\Gamma(\frac{i(3+\alpha)}{2})} \int T^{2i} d\!\mu^\alpha(T).
\]
When $z$ is inside the radius of convergence of the sum (i.e. $|z|<1$)
we can switch the 
integration and the summation and get the following expression for $f^\alpha$
\begin{eqnarray}
f^\alpha(z) &=& \int \sum_{j=0}^\infty \frac{T^{2j}z^{2j}}
{C^{2j} \Gamma(\frac{N(3+\alpha)}{2})} d\!\mu^\alpha(T)\\
&=& \int F^\alpha(zT) d\!\mu^\alpha(T) \label{eqn:int}.
\end{eqnarray}
We note a few simple facts. First notice that the function $f^\alpha(z)$ is a 
kind of generalized Laplace transform of the measure $d\!\mu^\alpha(T)$.
The quantity inside the integral, 
$F^\alpha(zT)=\sum\frac{ T^{2j}z^{2j}}{C^{2j}\Gamma(\frac{j(3+\alpha)}{2})}$ 
converges absolutely for all $z$ and thus 
$F^\alpha(zT)$ is an entire function of the complex variable $z$. 
Further we know that 
the integral must converge for $|z|<1$ and diverge 
for some $|z|>1$, since the original series converged in a circle of 
unit radius. We note that the entire function $F^\alpha(z)$ satisfies
\begin{eqnarray}
|F^\alpha(z)| &=& |\sum_{j=0}^\infty \frac{z^{2j}}
{C^{2j}\Gamma(i\frac{3+\alpha}{2})}| \\
&\le&\sum_{j=0}^\infty \frac{|z|^{2j}}{|C^{2j}\Gamma(j\frac{3+\alpha}{2})|} \\
&\le& F^\alpha(|z|),\label{eqn:bound} 
\end{eqnarray}
so the function $F^\alpha(z)$ grows fastest along the real axis. Thus we 
know that the integral in Equation (\ref{eqn:int}) converges for $-1<z<1$
and diverges for $z>1,z<-1$.    
Thus the problem of understanding the rate of decay of the tails of the 
probability measure $d\!\mu^\alpha(T)$ has been reduced to that of 
determining the rate of growth of the function $F(zt)$. There is a 
well-developed theory for studying the rate of growth of entire functions, 
the theory of entire functions of finite order. We recall only the basic 
facts here - the interested reader is referred to the texts of 
Ahlfors\cite{Ahlfors} and Rubel with Colliander\cite{RuCo}.  

The radial maximal function $M_F(r)$ of an entire function $F(z)$  
is defined to be the maximum of the absolute value of $F$ over 
a ball of radius $r$ centered on the origin:  
\[
M_F(r) = \max_{|z|=r}|F(z)| 
\]
The order $\rho$ of a function $F$ is defined to be 
\[
\rho = \limsup_{r\rightarrow\infty} \frac{\log_+\log_+ M_F(r)}{\log_+(r)},
\]
where $\log_+(x) = \max(0,\log(x))$, 
if this limit exists. It is easy to see from this definition that 
$F$ is of order $\rho$ means that $F$ grows asymptotically like 
$\exp(A(z) |z|^\rho)$ along the direction of maximum growth in the 
complex plane, where $A(z)$ grows more slowly than any power of $z$. 
A related notion is the type of a function of finite order. 
If $F$ is of order $\rho$ then the type $\tau$ is defined to be 
\[
\tau = \limsup_{r\rightarrow\infty} \frac{\log_+ M_F(r)}{r^\rho}
\]
when this limit exists. Again speaking very roughly the type $\tau$ 
gives the next order asymptotics: 
if $F$ is of order $\rho$ and type $\tau$ then $F$ grows like 
$B(z)\exp(\tau |z|^\rho)$, where $B(z)$ is subdominant to the 
exponential term. Note that by Equation (\ref{eqn:bound}) the function
$F^\alpha$ grows fastest along the real axis, and thus the maximal 
rate of growth in the complex plane is exactly the rate of growth 
along the real axis.   

There exist alternate characterizations of the order and type of a function 
in terms of the Taylor coefficients $A_n$ which are very useful 
for our purposes. These are given as follows:
\begin{eqnarray}
\rho &=& \limsup_{r\rightarrow\infty} \frac{\log_+\log_+ M_F(r)}{\log_+(r)}
= \limsup_{n\rightarrow\infty} \frac{n\log(n)}{-\log(|A_n|)} 
\label{eqn:order}\\
\tau &=& \limsup_{r\rightarrow\infty} \frac{\log_+M_F(r)}{r^\rho}
= \frac{1}{\rho e} \limsup_{n\rightarrow\infty} n|A_n|^{\rho/n}\label{eqn:type}.
\end{eqnarray}  
For the proofs we refer to the text of Rubel with Colliander\cite{RuCo}. 
 Using the expressions 
given in equations (\ref{eqn:order}) and (\ref{eqn:type}) 
we find that the order $\rho$ and type $\tau$ of $F^\alpha(z)$ are given 
by 
\begin{eqnarray*}
\rho^\alpha = \limsup_{n\rightarrow\infty} 
{\frac{2n \log(2n)}{\log(C^{2n}\Gamma(\frac{(3+\alpha)n}{2}))}} = 
{\frac{4}{3+\alpha}} \\
\tau^\alpha ={\frac{1}{\rho e}} \limsup_{n\rightarrow\infty} n |\Gamma(\frac{(3+\alpha)n}{2})|^{\frac{-\rho}{n}} = \frac{1}{C^\rho} \\
\end{eqnarray*}
Thus we know that $F^\alpha(zT)$ grows like 
$A(zT)\exp(C^{-\rho}|z|^{\frac{4}{3+\alpha}}|T|^{\frac{4}{3+\alpha}})$ 
along the real axis, where $A(zT)$ grows more slowly than 
$\exp(D|T|^\frac{4}{3+\alpha})$ for any $D$. Further we know that 
the integral
\[
\int F^\alpha(zT) d\!\mu^\alpha(T) 
\]
converges for $|z|<1$ and diverges for $z>1$ or $z<-1$, so to leading 
order the rate of decay of the measure $d\mu^\alpha(T)$ is given by 
$\exp(-|C|^{-4/(3+\alpha)}|T|^{4/(3+\alpha)})$. It is easy to check that 
as $\alpha \rightarrow -1$ this estimate becomes $\exp(-\frac{T^2}{4})$, 
recovering the normalized Gaussian. 
  
This result is probably best restated in terms of the cumulative 
distribution function, rather than the probability measure. If 
$P(T,T') = \int_T^{T'} d\mu(T)$, with $T'>T$, then it is easy to 
show that the above implies that
\begin{eqnarray*}
\lim_{T\rightarrow\infty} 
\exp(c |T|^{\frac{4}{3+\alpha}}) P(T,T') &=& 0 
\qquad  c < |C|^{\frac{-4}{3+\alpha}} \\
&=& \infty \qquad c > |C|^{\frac{-4}{3+\alpha}}
\end{eqnarray*}
 
\section{Interpretation and concluding remarks}

Physically the Majda model can be thought of as a model for the behavior 
of a passive scalar at small scales, when the scale of the flow field
is {\em much larger than the scale of the variations of the scalar.} Recall 
that the random scalar is given by 
\begin{eqnarray}
T(y) &=& \int |k|^{\frac{\alpha}{2}} \hat\phi_0(k) e^{2 \pi i k y} dW(k) \\
<T(y) T(y')> &=& \int |k|^\alpha |\hat\phi_0(k)|^2 e^{2 \pi i k(y-y')} dk.
\end{eqnarray}
In the limit as $\alpha$ approaches $-1$ there is an infrared divergence, 
so that the energy of the scalar is concentrated at larger and larger 
scales. In this case $\frac{4}{3+\alpha}\rightarrow 2$, so the 
normalized distribution function becomes Gaussian, as was originally 
observed by Majda. 

One important fact about this model which we would like to emphasize is 
that it predicts that higher derivatives of the advected scalar 
should be {\em increasingly intermittent}, a fact which was not 
strongly emphasized in previous work. Observe that due to the special nature 
of shear flows the scalar derivative $\partial T/\partial y$ satisfies the 
same equation as the scalar $T$ with {\em no additional terms!}.   
We further note that the initial condition for the derivative of the scalar 
is given by 
\begin{eqnarray}
\frac{\partial T}{\partial y} &=&
 \int 2 \pi i |k|^{\frac{\alpha}{2}} k \hat\phi_0(k) e^{2 \pi i k y}dW(k) \\
<\frac{\partial T}{\partial y}\frac{\partial T}{\partial y'}>
&=& 4 \pi^2 \int |k|^{\alpha+2} |\hat\phi_0(k)|^2 e^{2 \pi i k(y-y')} dk,
\end{eqnarray}
so the derivative of the scalar has a representation of the same 
form as the representation of the scalar itself, but with the exponent
$\alpha$ increased by two, and a slightly modified $\phi_0(k)$. Recall that 
the exponent $\alpha$ determines the amount of energy at the largest scales 
and thus the degree of intermittency, 
with the tails decaying as $\exp(-T^{4/(3+\alpha)})$.
Our calculation shows that increasing $\alpha$ 
increases the width of the tails of the probability distribution 
function, {\em implying that derivatives are more intermittent!} These 
predictions for the behavior of the tails of the scalar as compared 
with the scalar gradient are in extremely good agreement with 
experimental and numerical results. For instance our calculation
shows that if the scalar has 
exponent $\alpha=-1$, so that the probability distribution 
function of the scalar has Gaussian tails, then the derivative of the scalar 
has exponent $\alpha =1$, implying that the 
distribution of the derivative has {\em exponential tails.} This 
agrees quite well with the experiments of 
Van Atta and Thorddsen\cite{ThVanA}, as just one example, 
who observe that in turbulent thermally stratified flow that the pdf for 
the density has Gaussian tails, while the pdf of the density gradient has 
exponential tails.
Similarly if the scalar has exponent $\alpha=1$, so that the 
distribution of the scalar itself is exponential, then derivative of the 
scalar should have exponent $\frac{2}{3}$. This agrees with the
calculations of Chertkov, Falkovich and Kolokolov\cite{CFK},    
and Balkovsky and Falkovich\cite{BaFa} also predict exponential tails 
for the scalar and 
stretched exponential tails with exponent $\frac{2}{3}$ for the scalar 
gradient in the Batchelor regime. This also shows reasonably good 
agreement with the numerical experiments of Holzer and 
Siggia\cite{HoSig1,HoSig2}. In their experiments Holzer and Siggia find 
that a scalar with exponential tails has a gradient 
with stretched exponential tails. For large Peclet number the exponent 
of these stretched exponential tails is in the range of $.661-.563$. 

Of course one can eliminate $\alpha$ entirely, and one finds the following 
relationship between the distribution of the scalar and the scalar gradient 
within this model. If $T$ is distributed according to a stretched 
exponential pdf with exponent $\rho$, and the gradient $T_y$ according 
to a stretched exponential pdf with exponent $\rho'$, then $\rho,\rho'$ 
are related by 
\[
\frac{1}{2} + \frac{1}{\rho} = \frac{1}{\rho'}.
\]
It would be extremely interesting to check if this relationship, or some 
generalization of it, holds in greater generality than shear flows. The above 
numerical and experimental evidence suggest that this might not be an 
unreasonable hope.   

The distribution of the $x$, or cross-shear, derivatives can also 
be calculated using the same explicit representations derived by 
Majda. Calculations by the authors for deterministic initial  
data have shown that derivatives in the cross-shear direction have a 
distribution with the {\em same} asymptotic behavior as 
the scalar itself. This should be compared to and  contrasted with 
the papers of Son\cite{Son}, and Balkovsky and Fouxon\cite{BF}, which 
predict distributions with very broad tails (all of the higher moments 
diverge as $t \rightarrow \infty$) and which predicts the same 
distribution for derivatives of the scalar as for the scalar itself.  

We would also like to comment on the relationship between 
intermittency and the Lyapunov exponents of the underlying 
flow field. A number of papers have addressed the problem of 
intermittency in the large Peclet number limit by attempting to 
relate broader than Gaussian tails to the Lyapnuov exponents 
of the flow field\cite{CFKL}. It is worth noting that a shear flow does not 
possess a positive Lyapunov exponent, but as we have shown here 
a shear flow can generate exponential and stretched exponential 
tails in the passive scalar. 
This shows that chaotic behavior in the underlying flow, while 
probably an important effect in realistic flows, is not necessary
for the generation of broad tails and intermittency.

Finally we would like to comment on the rate of approach to the limiting 
measure in time. The results presented here analyze the infinite time 
limit of the measure. As mentioned earlier the convergence to this limiting
measure is expected to be highly non-uniform. A preliminary calculation 
by the authors for a special choice of the cut-off function $\hat\phi_0(k)$ 
suggests that for large but finite times the pdf looks like the pdf for the 
infinite time problem in some core region, with Gaussian tails 
outside this core region. As time increases the size of this core 
region demonstrating non-Gaussian statistics grows, and the Gaussian 
tails get pushed out to infinity. We believe this same picture to 
hold for any choice of the cut-off function $\hat\phi_0(k)$, but 
more work is needed to establish this fact. 
   
{\bf Acknowledgements:} Jared C. Bronski would like to acknowledge 
support from the National Science Foundation under grant DMS-9972869. 
Richard M. McLaughlin would like to acknowledge support from 
NSF Career Grant DMS-97019242, 
and would like to thank L. Kadanoff and the James Franck Institute
for support during the writing of this paper, and Raymond T. Pierrehumbert 
for several useful conversations.    
The authors would like to thank Misha Chertkov, Leo Kadanoff and Kenneth 
T-R. McLaughlin for several conversations, and Pete Kramer for an extremely 
thorough reading of the original manuscript.

\end{document}